\documentclass[fleqn,usenatbib]{mnras}

\usepackage{newtxtext,newtxmath}
\usepackage{subcaption}
\usepackage{physics}

\usepackage[T1]{fontenc}

\DeclareRobustCommand{\VAN}[3]{#2}
\let\VANthebibliography\thebibliography
\def\thebibliography{\DeclareRobustCommand{\VAN}[3]{##3}\VANthebibliography}

\usepackage{graphicx}	
\usepackage{amsmath}	

\title[Rotating filament]{A 15\,Mpc rotating galaxy filament at redshift $z=0.032$}

\author[M. N. Tudorache et al.]{
Madalina N. Tudorache,$^{1, 2}$\thanks{e-mail: madalina.tudorache@ast.cam.ac.uk}
S. L. Jung,$^{2}$\thanks{e-mail: lyla.jung@physics.ox.ac.uk}
M. J. Jarvis,$^{2,3}$
I. Heywood,$^{2,4,5}$
A. A. Ponomareva,$^{6,2}$
\newauthor
A. Varasteanu,$^{2}$
N. Maddox,$^{7}$
T. Yasin,$^{2}$
M. Glowacki$^{8,9}$
\\
$^{1}${Institute of Astronomy, University of Cambridge, Madingley Road, Cambridge CB3 0HA, UK} \\
$^{2}${Astrophysics, Department of Physics, University of Oxford, Keble Road, Oxford OX1 3RH, UK} \\
$^{3}${Department of Physics and Astronomy, University of the Western Cape, Robert Sobukwe Road, 7535 Bellville, Cape Town, South Africa} \\
$^{4}${Department of Physics and Electronics, Rhodes University, PO Box 94, Makhanda, 6140, South Africa} \\
$^{5}${South African Radio Astronomy Observatory, 2 Fir Street, Black River Park, Observatory, Cape Town 7925, South Africa} \\
$^{6}${Centre for Astrophysics Research, School of Physics, Astronomy and Mathematics, University of Hertfordshire, College Lane} \\
$^{7}${School of Physics, H.H. Wills Physics Laboratory, Tyndall Avenue, University of Bristol, Bristol, BS8 1TL, UK} \\
$^{8}${Institute for Astronomy, University of Edinburgh, Royal Observatory, Edinburgh, EH9 3HJ, United Kingdom}\\
$^{9}${The Inter-University Institute for Data Intensive Astronomy (IDIA), Department of Astronomy, University of Cape Town, Private Bag X3, Rondebosch 7701, South Africa}}

\date{Accepted XXX. Received YYY; in original form ZZZ}

\pubyear{\the\year{}}

\begin{document}
\label{firstpage}
\pagerange{\pageref{firstpage}--\pageref{lastpage}}
\maketitle

\begin{abstract}
Understanding the cold atomic hydrogen gas (H{\sc i}) within cosmic filaments has the potential to pin down the relationship between the low density gas in the cosmic web and how the galaxies that lie within it grow using this material. We report the discovery of a cosmic filament using
$14$ H{\sc i}-selected galaxies that form a very thin elongated structure of $1.7$ Mpc. These galaxies are embedded within a much larger cosmic web filament, traced by optical galaxies, that spans at least $\sim 15$~Mpc. We find that the spin axes of the H{\sc i} galaxies are significantly more strongly aligned with the cosmic web filament ($\langle\lvert \cos \psi \rvert\rangle = 0.64 \pm 0.05$) than cosmological simulations predict, with the optically-selected galaxies showing alignment to a lesser degree ($\langle\lvert \cos \psi \rvert\rangle = 0.55 \pm 0.05$). This structure demonstrates that within the cosmic filament, the angular momentum of galaxies is closely connected to the large-scale filamentary structure. We also find strong evidence that the galaxies are orbiting around the spine of the filament, making this one of the largest rotating structures discovered thus far, and from which we can infer that there is transfer of angular momentum from the filament to the individual galaxies. The abundance of H{\sc i} galaxies along the filament and the low dynamical temperature of the galaxies within the filament indicates that this filament is at an early evolutionary stage where the imprint of cosmic matter flow on galaxies has been preserved over cosmic time.

\end{abstract}

\begin{keywords}
cosmology: large-scale structure of Universe -- cosmology: dark matter -- galaxies: evolution
\end{keywords}



\section{Introduction}

The Universe, on the largest scale, exhibits a network-like distribution of matter - known as the "cosmic web" \citep{bond-1996}. 
According to Zel'dovich's model of non-linear growth, primordial density perturbations are amplified by gravitational instabilities \citep{zeldovich-1970}. Cosmological-volume $N$-body simulations confirm the theory, showing that cold dark matter (CDM) forms the cosmic web, comprised of clusters, walls, voids and filaments \citep{springel_2005}.
Cosmic filaments form where walls of matter intersect and collapse. Tidal Torque Theory \citep{Doroshkevich1970,White_1984,CatelanTheuns} predicts that asymmetry in this process results in tidal forces that can produce a torque, resulting in rotation of filaments as they form, with the filaments subsequently acting as highways along which matter can flow towards the intersection with other filaments, known as nodes \citep{Pichon_2011, Trowland_2013, Codis_2015, Laigle_2015, Xia_2021}.  

Although there are ongoing attempts to trace the dark component of cosmic filaments via gravitational weak lensing analyses \citep{Kim_2024} and the diffuse gas component via atomic hydrogen (H{\sc i}) emission \citep{Kooistra_2017,Tramonte_2019}, much of our knowledge about the cosmic large-scale structure relies on the distribution of galaxies as tracers. The distribution of galaxies found in early studies \citep{Davis1982, deLapparent-1986} represent our earliest observational evidence for the cosmic web.

On the individual galaxy scale, galaxies in cosmic filament environments should inherit spin from the dynamics of the neighbouring matter in filaments. Theory and numerical simulations have shown that high-mass and low-mass galaxies respond differently to the anisotropic filamentary structure \citep{Aragon-Calvo_2007, Codis_2012, Libeskind_2013, Dubois_2014, Wang_2017}. 
In these simulations, low-mass systems, predominantly disk-like spiral and dwarf galaxies, tend to have their spin axes parallel to the orientation of their nearest filament. More massive systems, predominantly elliptical galaxies, tend to have their spin axes perpendicular to the orientation of the filament, which may subsequently influence how much energy is deposited in a given direction from active galactic nuclei \citep[e.g.][]{Jung_2025}.

This mass-dependent spin alignment can be explained in the hierarchical galaxy formation framework. The first galaxies and dark matter halos form within a vorticity cell of filaments that is aligned with the filament orientation \citep{Codis_2015, Laigle_2015}. As the Universe evolves, the filaments themselves begin to collapse and galaxies and dark matter halos traverse towards the nodes at the intersection of the filaments.
Some of these halos and galaxies merge, disrupting their spin-axis alignment with the filament. As they become more massive merged systems, their spin axis is reoriented perpendicular to the filament as their major axis becomes aligned with the direction of flow along the filament, due to gravitational forces acting on the dark matter halo and galaxy. 
Recent observational evidence for a bulk rotation of galaxies around the filament spines has been revealed by \citet{Wang_2021}. By stacking thousands of cosmic filaments they find that galaxies rotate around the filament spine, making filaments some of the largest rotating structures in the Universe, suggesting that angular momentum can be generated on very large scales.



Large filamentary structures have only recently been discovered using H{\sc i} observations. \cite{Arabsalmani_2025} identified a straight narrow filament of galaxies that extends about $5\,\rm Mpc$ at a redshift of $z \sim 0.0365$. \cite{Lawrie_2025} find, also using MeerKAT, a filamentary-like overdensity of H{\sc i} galaxies at a redshift of $z \sim 0.0395$.

In this paper, we present the detection of a string of H{\sc i} galaxies first discovered in the recent Data Release 1 of the  H{\sc i} emission line component of the MeerKAT International Giga-Hertz Tiered Extragalactic Exploration (MIGHTEE-H{\sc i}) survey \citep{jarvis2016, Maddox_2021, Heywood_2024}. These galaxies map on to a much larger single cosmic filament defined by optically-selected galaxies with precise spectroscopic redshifts from the Dark Energy Spectroscopic Instrument survey \citep{Dey2019}. 
The structure of this paper is organised as follows. Section \ref{sec:methods} introduces the MIGHTEE Survey, as well as the methods employed to compute the cosmic web and the spin of the H{\sc i} galaxies. Section \ref{sec:results} presents our results and Section \ref{sec:discussion} discusses the results obtained. The summary and conclusions are presented in Section \ref{sec:conclusions}. 

We assume $\Lambda$CDM cosmology with $H_0 = 70$ km~s$^{-1}$~Mpc$^{-1}$, $\Omega_{\rm M} = 0.3$ and $\Omega_\Lambda = 0.7$.

\section{Methods}
\label{sec:methods}

\subsection{MIGHTEE-H{\sc i} data}
\label{subsec:data}

The MIGHTEE survey is one of the eight Large Survey Projects (LSPs) which are being undertaken by MeerKAT \citep{meerkat}. MeerKAT consists of an array of 64 offset-Gregorian dishes, where each dish consists of a main reflector with a diameter of $13.5$\,m and a sub-reflector with a diameter of $3.8$\,m. MeerKAT's three band receivers, UHF–band ($580 < \nu < 1015 $\,MHz), L--band ($900 < \nu < 1670$\,MHz) and S–band ($1750 < \nu < 3500$\,MHz) all collect data in spectral mode. The MIGHTEE survey has three major components: radio continuum \citep{Heywood2021}, polarisation \citep{Taylor_2024} and spectral line \citep{Maddox_2021}. MIGHTEE-H{\sc i} \citep{Maddox_2021} is the H{\sc i} emission part of the MIGHTEE survey. Data Release 1 \citep{Heywood_2024} is comprised of 15 mosaicked pointing with $94.2$~h of integration time. 
The data products were by using a parallelised 
\textsc{CASA}\footnote{\url{http://casa.nrao.edu}}-based \citep{casa} pipeline with standard routines (i.e. flagging, delay, bandpass, and complex gain calibration, \citealt{Heywood_2024}). Visibility flagging is conducted using the \textsc{tricolour} package \citep{Hugo_2022}. Each sub-band undergoes imaging via the \textsc{wsclean} software \citep{Offringa_2014}, with a pointing-specific mask derived from deep MIGHTEE continuum images \citep{Heywood_2022}. The spectral clean component model is smoothed using the \textsc{smops}\footnote{\url{https://github.com/Mulan-94/smops}} tool for spectral smoothness. After inversion into the visibility domain, (phase+delay) self-calibration and simultaneous subtraction of the continuum model occur using the \textsc{cubical} package \citep{Kenyon_2018}. Pointings are then imaged per-channel with three robustness parameters (0.0, 0.5, and 1.0) using \textsc{wsclean} \citep{Briggs_1995}, and deconvolution masks are generated with a custom Python tool \citep[see ][]{Heywood_2024}. Imaging is repeated within masked regions, and resulting cubes are homogenised to a common angular resolution per channel using a custom \textsc{Python} code and the \textsc{pypher} package \citep{Boucaud_2016}. These homogenised images are primary beam corrected with the \textsc{katbeam}\footnote{\url{https://github.com/ska-sa/katbeam}} library and linearly mosaicked with variance weighting using the \textsc{montage}\footnote{\url{http://montage.ipac.caltech.edu/}} toolkit. Finally, image-plane continuum subtraction is performed along each sightline through the resulting cubes. 

For this work, we use the spectral line information in the L2 band ($1310-1420$\,MHz) with $32768$ channels with a channel width of $26.5$\,kHz, which corresponds to $5.5$\,km\,s$^{-1}$ at $z=0$.

The H{\sc i} mass of each galaxy is calculated from the integrated flux $S$, as:

\begin{equation}
\left(\frac{M_{\mathrm{HI}}}{\mathrm{M_{\odot}}}\right)=\frac{2.356 \times 10^5}{1+z}\left(\frac{D_L}{\mathrm{Mpc}}\right)^2\left(\frac{S}{\mathrm{Jy}~\mathrm{km} \mathrm{s}^{-1}}\right),
\end{equation}
where $D_L$ is the cosmological luminosity distance to the source, $S$ is the integrated H{\sc i} flux density, calculated from the moment-0 (integrated intensity over the spectral line) H{\sc i} maps.

\begin{table*}
\centering
\begin{tabular}{ccccccccc}
\hline
 ID &  RA (h:m:s) &   Dec (d:m:s) &  Velocity (km/s) & z &  $\log_{10}{M_{\mathrm{HI}}/M_{\odot}}$ &  PA (deg) & $i$ (deg) & $\log_{10}{M_{\ast}/M_{\odot}}$ \\
 \hline\hline
 1 & 9:57:13 & 2:08:16 & 9530 & 0.032 & 8.4 & 25.6 & 52.9 & -  \\
 2 & 9:57:13 & 2:07:08 & 9320 & 0.031 & 8.5 & 241.1 & 24.6 & $9.1 \pm 0.3$ \\
 3 & 9:57:44 & 2:00:03 & 9380 & 0.031 & 8.5 & 320.1 & 47.2 & $7.4 \pm 0.3$ \\
 4 & 9:57:27 & 1:59:06 & 9530 & 0.032 & 8.3 & 58.6 & 8.2 & $8.1 \pm 0.4$  \\
 5 & 9:58:02 & 1:57:14 & 9340 & 0.031 & 8.7 & 131.4 & 70.8 & $8.5 \pm 0.3$ \\
 6 & 9:57:20 & 1:55:08 & 9450 & 0.032 & 9.6 & 55.5 & 37.1 & $10.2 \pm 0.5$ \\
 7 & 9:57:12 & 1:54:57 & 9375 & 0.031 & 8.4 & 110.0 & 9.7 & $8.6 \pm 0.4$ \\
 8 & 9:57:27 & 1:52:22 & 9700 & 0.032 & 8.9 & 246.8 & 17.9 & $9.1 \pm 0.4$ \\
 9 & 9:57:53 & 1:48:18 & 9300 & 0.031 & 8.1 & 320.0 & 6.3 & $7.8 \pm 0.3$ \\
 10 & 9:57:32 & 1:40:33 & 9610 & 0.032 & 8.2 & 216.9 & 7.1 & $9.5 \pm 0.4$\\
 11 & 9:57:33 & 1:39:36 & 9700 & 0.032 & 9.1 & 7.0 & 20.5 & $9.3 \pm 0.3$ \\
 12 & 9:57:36 & 1:35:09 & 9460 & 0.032 & 9.0 & 135.0 & 19.3 & $9.1 \pm 0.4$ \\
 13 & 9:58:06 & 1:30:50 & 9570 & 0.032 & 9.1 & 70.8 & 20.1 & $8.7 \pm 0.4$ \\
 14 & 9:57:39 & 1:24:03 & 9650 & 0.032 & 9.1 & 43.0 & 21.9 & $9.1 \pm 0.5$\\
\hline
\end{tabular}
\caption{Properties of the H{\sc i} galaxy sample. The H{\sc i} properties and their uncertainties are measured from the MIGHTEE data, whilst the stellar masses and their uncertainties are obtained from crossmatching with the SDSS + DESI sample. For the H{\sc i} masses and the PAs, the uncertainties are of order $10\%$, whilst for the inclination, the uncertainties are of order $\sim 5^{\circ}$ \citep{Varasteanu_2025}.}
\label{tab:properties}
\end{table*}

\subsection{Cosmic Web characterisation}
\label{subsec:cosmic-web}
We use \textsc{Disperse} \citep{sousbie} to determine the skeleton of the cosmic web based on the distribution of galaxies from SDSS DR17 \citep{Abdurrouf_2022} and DESI Legacy Data \citep{Dey2019, DESI_2023} with a stellar mass cut of $M_{\ast} > 10^{9} \rm M_{\odot}$, to ensure that we isolate the main structure. \textsc{Disperse} is a topological algorithm based on discrete Morse theory \citep{milnor1963morse} that computes the skeleton of the cosmic web using Delaunay Tessellation Field Estimator \citep{dtfe}. The algorithm uses the spatial position and redshift of galaxies to construct a filament network which is described by segments which connect the maxima, minima and saddle points of the density field.

The standard procedure in observational studies is to use the mirror boundary conditions in \textsc{Disperse} \citep{bird2019chiles, tudorache2022}.
When computing a filament, we adopt a 5$\sigma$ significance level, to ensure that the filament is robust and that larger structures are identified without missing some of the relevant finer structures.
In order to calculate the distances between the filament obtained with \textsc{Disperse} and the galaxies, we crossmatch the galaxies with the mid-point of each segment generated by \textsc{Disperse} to find the appropriate segment, and then we calculate the perpendicular distance to that segment.

\subsection{Calculating the Spin-filament alignment}
\label{subsec:spin-filament-calc}
By using a thin-disk approximation \citep{Lee_2007}, the spin unit vector of a galaxy can be characterised in local spherical coordinates as:
\begin{align}
    \hat{L}_{r} &=\cos i \\
    \hat{L}_{\theta} &= \sin i \sin \mathrm{PA}   \label{eq:hi-spin-sph2}\\
    \hat{L}_{\phi} &= \sin i \cos \mathrm{PA}
    \label{eq:hi-spin-sph3}
\end{align}
where PA is the position angle and $i$ is the inclination angle of the galaxy. To determine the inclination for the galaxies, 
we use {\sc photutils}\footnote{\url{https://zenodo.org/records/10967176}} to measure the ellipticity and position angle of the galaxy on the sky, fixing the centre of the galaxy and fitting isophotes to the $g$-band image, which traces the extended disk emission in the galaxies. Then, we define the inclination as:
\begin{equation}
    \cos(i)=b/a,
\end{equation}
where $b$ and $a$ are the minor and major axes of the fitted ellipse.

The unit spin vector is converted into Cartesian coordinates, with the spherical vector related to the Cartesian vector by:
\begin{equation}
\left[\begin{array}{c}
\hat{L}_{x} \\
\hat{L}_{y} \\
\hat{L}_{z}
\end{array}\right]=\left[\begin{array}{ccc}
\sin \alpha \cos \beta & \cos \alpha \cos \beta & -\sin \beta \\
\sin \alpha \sin \beta & \cos \alpha \sin \beta & \cos \beta \\
\cos \alpha & -\sin \alpha & 0
\end{array}\right]\left[\begin{array}{c}
\hat{L}_{r} \\
\hat{L}_{\theta} \\
\hat{L}_{\phi}
\end{array}\right] ,
\label{eq:hi-spin-xyz}
\end{equation}
where $\alpha=\pi / 2-\mathrm{Dec}$ and $\beta=\mathrm{RA}$, with Dec and RA corresponding to declination and right ascension, respectively.
As there is a sign ambiguity which arises in $\hat{L}_{r}$ \citep{Trujillo_2006}, we follow past work \citep{Lee_2007, Kraljic_2021, tudorache2022, Barsanti_2022, Barsanti_2023}, and take the positive sign in $\hat{L}_{r}$. Because the 3D spin vector depends non-linearly on both position angle and inclination, and we cannot measure the direction of the tilt of the galaxy (i.e. if the inclination is tilted towards or away from us) this will introduce a degeneracy in the spin-filament alignment signal, as the filament is defined in three dimensions. In practice, this translates to a $\pm \pi$ uncertainty in the spin-axis/minor axis position angle, which propagates through to the uncertainty in the dot product between the filament vector and the galaxy spin vector due to the $\sin \rm PA$ term in Eqn.~\ref{eq:hi-spin-sph2}.  This will become relevant in Section~\ref{subsec:spin-filament}, where we discuss how this uncertainty affects the strength of the signal.

We also determine the filament vector, which is defined by a starting point $f_1(\mathrm{RA}_1, \mathrm{Dec}_1, \mathrm{z}_1)$ and an end point $f_2(\mathrm{RA}_2, \mathrm{Dec}_2, \mathrm{z}_2)$.
\begin{figure}
\centering
\includegraphics[width=0.95\linewidth]{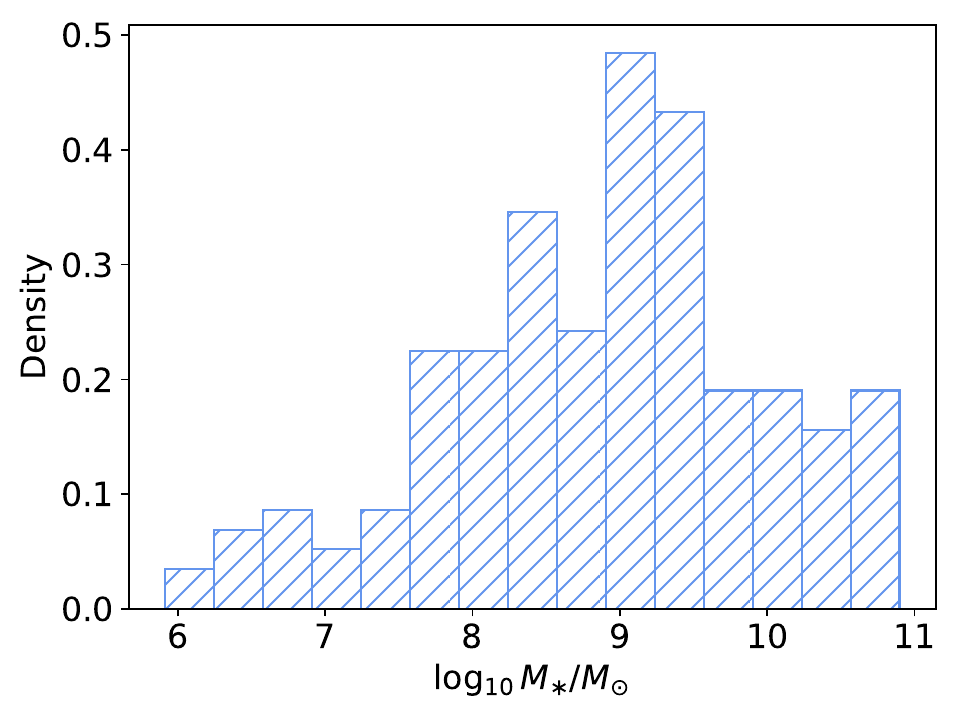}
\caption{Normalised histogram of the optical galaxies from DESI in the redshift range of $ 0.03 < $ z $ < 0.034$ as a function of stellar mass.}
\label{fig:galaxies-mass}
\end{figure}
These points are all generated using the skeleton from \textsc{Disperse}. To calculate the spherical components of the filament vector we can write:

\begin{align}
    f_{\mathrm{RA}} &= f_2(\mathrm{RA}_2) - f_1(\mathrm{RA}_1) ,\\
    f_{\mathrm{Dec}} &= f_2(\mathrm{Dec}_2) - f_1(\mathrm{Dec}_1) ,\\
    f_{\mathrm{z}} &= f_2(\mathrm{z}_{2}) - f_1(\mathrm{z}_{1}).
    \label{eq:filament}
\end{align}

The filament vector is then converted into Cartesian coordinates in order to compute the dot product between the spin vector of the galaxy $\mathbf{L}$ and the filament vector $\mathbf{f}$. To find the cosine of the angle between the galaxy spin vector and the filament vector $\psi$, we divide the dot product by the modulus of the filament vector, as the spin vector is already normalised to 1:
\begin{equation}
    \cos \psi = \frac{f_x \cdot \hat{L}_x + f_y \cdot \hat{L}_y + f_z \cdot \hat{L}_z}{|\mathbf{f}|},
\end{equation}
where $f_x$, $f_y$, $f_z$ are the Cartesian components of the filament vector $\mathbf{f}$ and $\hat{L}_x$, $\hat{L}_y$, $\hat{L}_z$ are defined as before. We refer to Figure~3 in \citet{tudorache2022} for a representative schematic of both the spin vector and the spin-filament angle.

\section{Results}
\label{sec:results}

\subsection{Discovery and characterisation of a 15~Mpc cosmic filament}
\begin{figure*}
\centering
\includegraphics[width=0.85\linewidth]{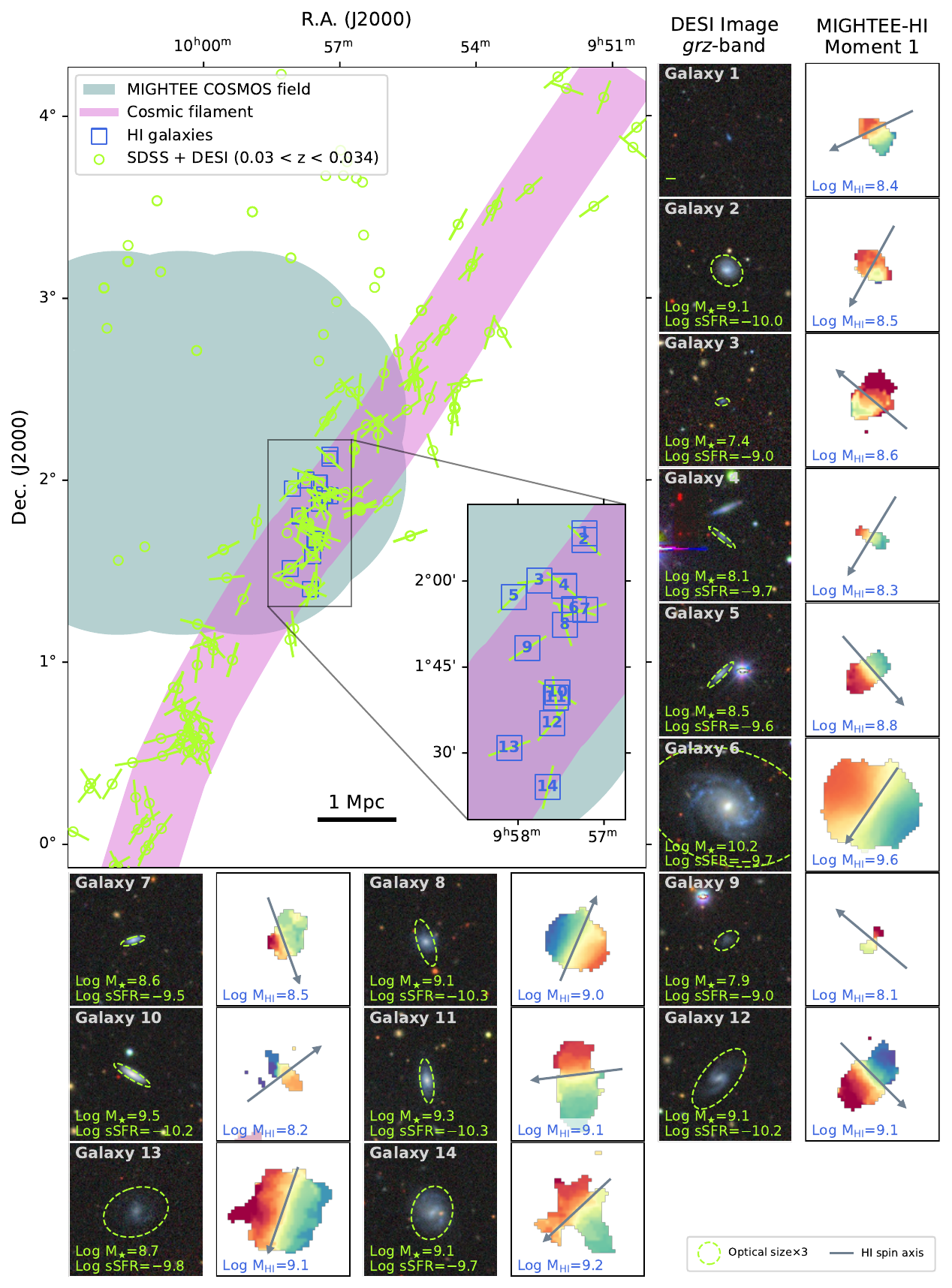}
\caption{Top left: the on-sky distribution of H{\sc i} galaxies (blue squares), SDSS and DESI optical galaxies (green circles and line, depending on the availability of optical PA measurements), and the cosmic filament. The MIGHTEE COSMOS footprint is shown in a grey block. Other panels show the DESI multi-band cutout image and the H{\sc i} moment-1 map of each H{\sc i}-selected galaxy. The size of the images and moment maps is fixed to 16 arcsec across each panel. 
The green dashed ellipse in each DESI image panel shows the ellipticity and the size (tripled for visual purposes) of the optical counterpart of each H{\sc i} galaxy. Note that Galaxy 1 does not have an optical DESI counterpart. The grey colour arrow in the moment-1 map is the H{\sc i} spin axis.}
\label{fig:poster}
\end{figure*}

We first identified an alignment of 14 H{\sc i} galaxies within the recently released MIGHTEE-H{\sc i} survey \citep{Heywood_2024} over the well studied COSMOS field \citep{scoville_2007}.
We present their on-sky positions and the H{\sc i} moment-1 map in Figure~\ref{fig:poster}, along with the multi-colour image of their optical counterparts from the Dark Energy Spectroscopic Instrument Legacy Imaging \citep{Dey2019}. The full coverage of the MIGHTEE-H{\sc i} COSMOS field is shown as a grey block in the main panel.
The 14 galaxies all lie within a very small range in recessional velocity between 9230 -- 9700~km~s$^{-1}$ (with a velocity dispersion of $\sim 140$~km/s) and form a linear structure at around 30~degree west of north on the celestial sphere, with a length of $ \sim 1.7$~Mpc and a width of $\sim 36$~kpc.

The linear structure is situated towards the corner of the current coverage of the H{\sc i} data cube and therefore it was unclear whether these data alone fully captured the whole structure. We therefore used optical spectroscopic data from DESI \citep{DESI_2023, Abdurrouf_2022} and SDSS which covers a wider area around the field, to explore the full extent of the structure. We found an additional 283 galaxies within the redshift range $0.03<z<0.034$, corresponding to 9065 -- 9865~km~s$^{-1}$. These galaxies span a stellar mass range of $10^{6}$\,M$_{\odot} < M_{\star} < 10^{11}$\,M$_{\odot}$, as can be seen in Figure \ref{fig:galaxies-mass}. We then use the same redshift slice, which contains $154$ spectroscopically confirmed galaxies with a stellar mass $M_{\star} > 10^{9}$\,M$_{\odot}$ with the \textsc{Disperse} setup described in Section {\ref{subsec:cosmic-web}} in order to define a filament.

The total length of the identified filament is $15.4\,\rm Mpc$ (sky-projected length of $12.4\,\rm Mpc$), with an average inclination of $37^{\circ}$ from the sky plane.
The pink-colour shaded line in Figure~\ref{fig:poster} shows the identified filament location and the galaxies within the field. 

\begin{figure}
  \centering
  \includegraphics[width=1\linewidth]{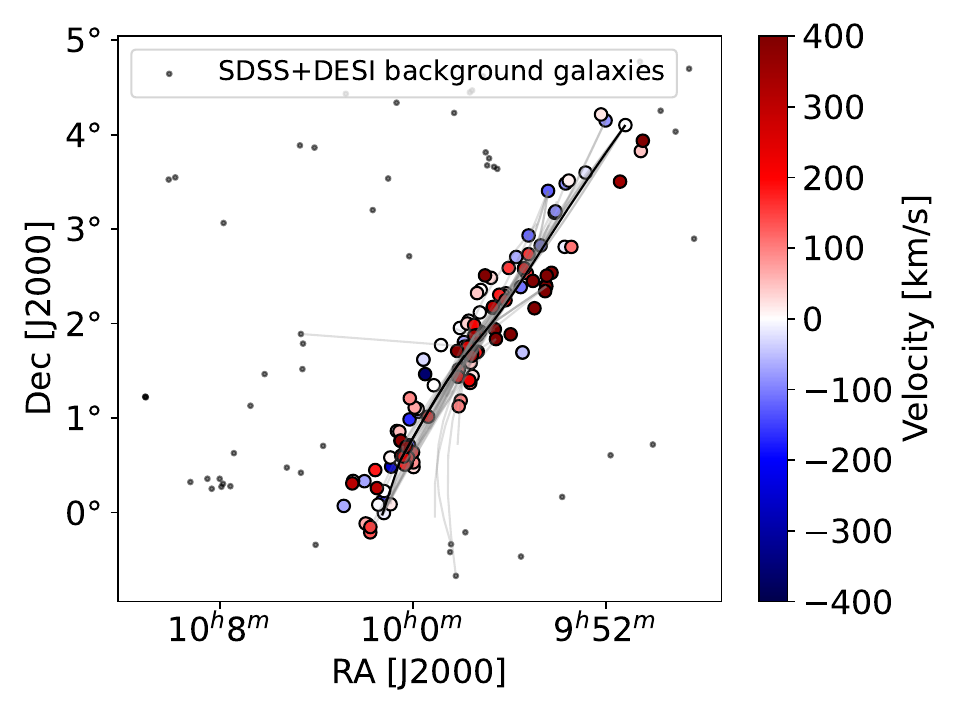}
  \caption{The rotation velocity of the filament galaxies which have a projected distance to filament within 1~Mpc. The colour bar shows the relative velocity to the filament. The black dots denote the non-filament galaxies in the DESI observations in the redshift interval $0.03 <  z < 0.034$. The thin gray lines represent the bootstrapped filament iterations.}
  \label{fig:rotation}
\end{figure}

It is non-trivial to calculate the accurate uncertainties on the position of the individual galaxies with respect to the filament due to the nature of the filament-finding algorithm. Hence, we use a jack-knife sampling of the filament by randomly omitting 5 per cent of the optical galaxies and re-determining a new filament $100$ times. As can be seen in Figure \ref{fig:rotation}, the main spine persists across all these iterations.

\subsection{Characteristic boundary of the filament}\label{sec:numdens}

The geometry of a cosmic filament in simulations is usually modelled as a cylinder, which can be defined by its length and its radius \citep{Galarraga-Espinosa_2020, Zakharova_2023, Galarraga-Espinosa_2024}. A density profile can be fit to its dark matter distribution in order to obtain the parameters which describe this geometry - leading to different results depending on the type of filament.  However, computing its radius (or its 'thickness') in observations is not straightforward, since we do not have access to dark matter densities and therefore have to rely on galaxies to calculate this quantity.
However, at low redshifts, most studies agree that filaments should have radii between $1$ to $2$~Mpc \citep{Bond_2010, Galarraga-Espinosa_2024, Wang_2024, Yang_2025}.

\begin{figure}
\centering
\includegraphics[width=0.95\linewidth]{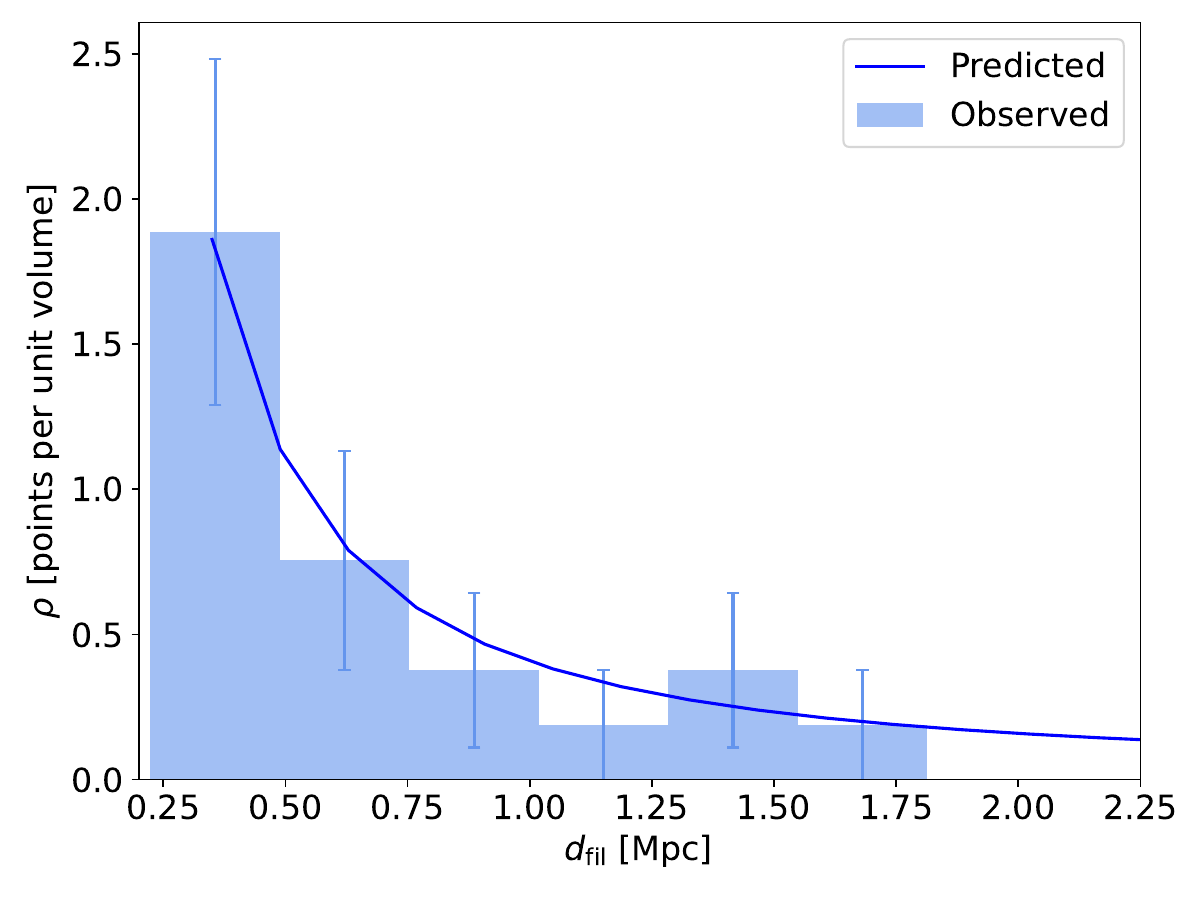}
\caption{Number density of galaxies as a function of distance-to-filament. The blue line represents the modelled data assuming a cylinder of uniform density (see text).}
\label{fig:filament-density}
\end{figure}

Here, we also model the filament as a cylinder and count the number of galaxies as a function of radius, from which we obtain a number density. 
We extend the measurement out to the radius at which the the number density reaches a constant. The galaxy number density as a function of radius away from the filament spine is shown in Figure~\ref{fig:filament-density}. As can be seen, most of the points are situated within $\sim 1$~Mpc of the spine of the filament. The error bars on the bins are calculated as Poisson errors. As a toy model, we assume that the galaxies are distributed uniformly within a cylinder, and that we are observing the cylinder as a projection from its base. As such, the number density should be normalised by the area of the arc of the base, $a$, defined as:
\begin{equation}
a=R^2 \arccos \left(1-\frac{h}{R}\right)-(R-h) \sqrt{R^2-(R-h)^2},
\end{equation}
where $R$ is the radius of the cylinder (filament) and $h$ is the varying distance from the spine. As can be seen in Figure \ref{fig:filament-density}, this model follows our number density distribution well within the error bars.

We also fit a polynomial function to the number density as a function of distance to spine of the filament to find its boundary, following \citet{Wang_2024}. We use a 6th order polynomial to fit the distribution, however, we find that a 6th order overfits the data at the end points. We therefore, also use a 5th order polynomial for our sparsely sampled data on just a single filament.

\begin{figure}
\centering
\includegraphics[width=0.8\linewidth]{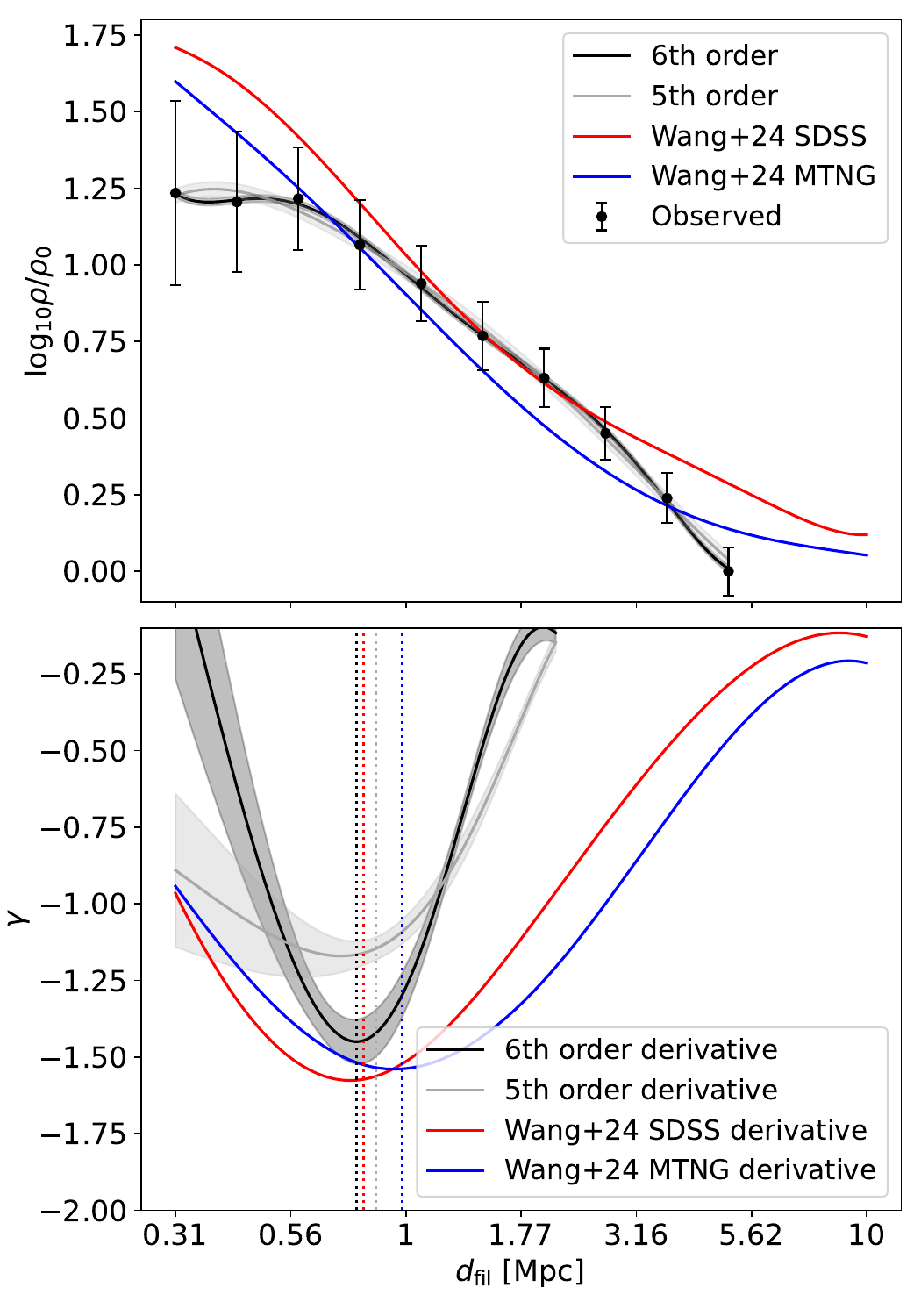}
\caption{{\it (top)} Number density of galaxies as a function of distance-to-filament, with Poissonian uncertainties. The black (grey) line denotes the 6th (5th) order  polynomial fits to the data, with residual uncertainties denoted by shaded regions. {\it (bottom)} The differential of the polynomial fits as a function of distance-to-filament. The vertical dotted lines show the minima (black for the 5th order derivative, gray for the 4th order derivative), corresponding to the characteristic radius of the filament.}
\label{fig:filament-density-fit}
\end{figure}

We use the derivative of the fitted polynomial, $\gamma = \dd \log_{10}{\rho}/\dd\log_{10}{d_{\text{fil}}}$, as a function of the distance-to-filament, following \cite{Diemer_2014} who use a similar method to study radial density gradients of dark matter halos in simulations. In a similar way, $\gamma$ can be used to define the variation in the number density as it departs from the spine of the filament (i.e. how fast it decreases). Its minimum represents the effective radius of the filament (but with some caveats, see \citealt{Wang_2024}). We show $\gamma$ as a function of distance from the filament spine in the bottom-panel of Figure~\ref{fig:filament-density-fit}. We find a minimum value for $\gamma$ at $R = 0.78 \pm 0.07$~Mpc for the 6th order polynomial fit and $R = 0.86 \pm 0.04$~Mpc for the 5th order polynomial. Both of these values for the effective radius of the filament fall within the range of those measured from the observational study using the stacked filaments from SDSS ($R = 0.81$~Mpc) and from the stacked filaments derived from the Millenium-TNG simulation ($R = 0.98$~Mpc). 
We find a minimum $\gamma \simeq -1.43$ for the 6th order polynomial ($\gamma \simeq -1.2$ for the 5th order polynomial) - which is slightly higher than the value of $\gamma \sim -1.6$ found in \cite{Wang_2024}, but significantly less steep than the results usually found for haloes (see \citealt{Diemer_2014}). 
This suggests that the density profile for this filament has a shallower profile than for the general filament population identified both in observations and simulations, from which we can infer that it is likely to be younger and has had less time for matter to collapse. Furthermore, the presence of angular momentum, which we discuss in Section \ref{subsec:rotation}, could also prevent the filament from collapsing further. 

\subsection{Galaxy Spin Axis -- Filament alignment}
\label{subsec:spin-filament}

\begin{figure}
  \centering
  \includegraphics[width=1\linewidth]{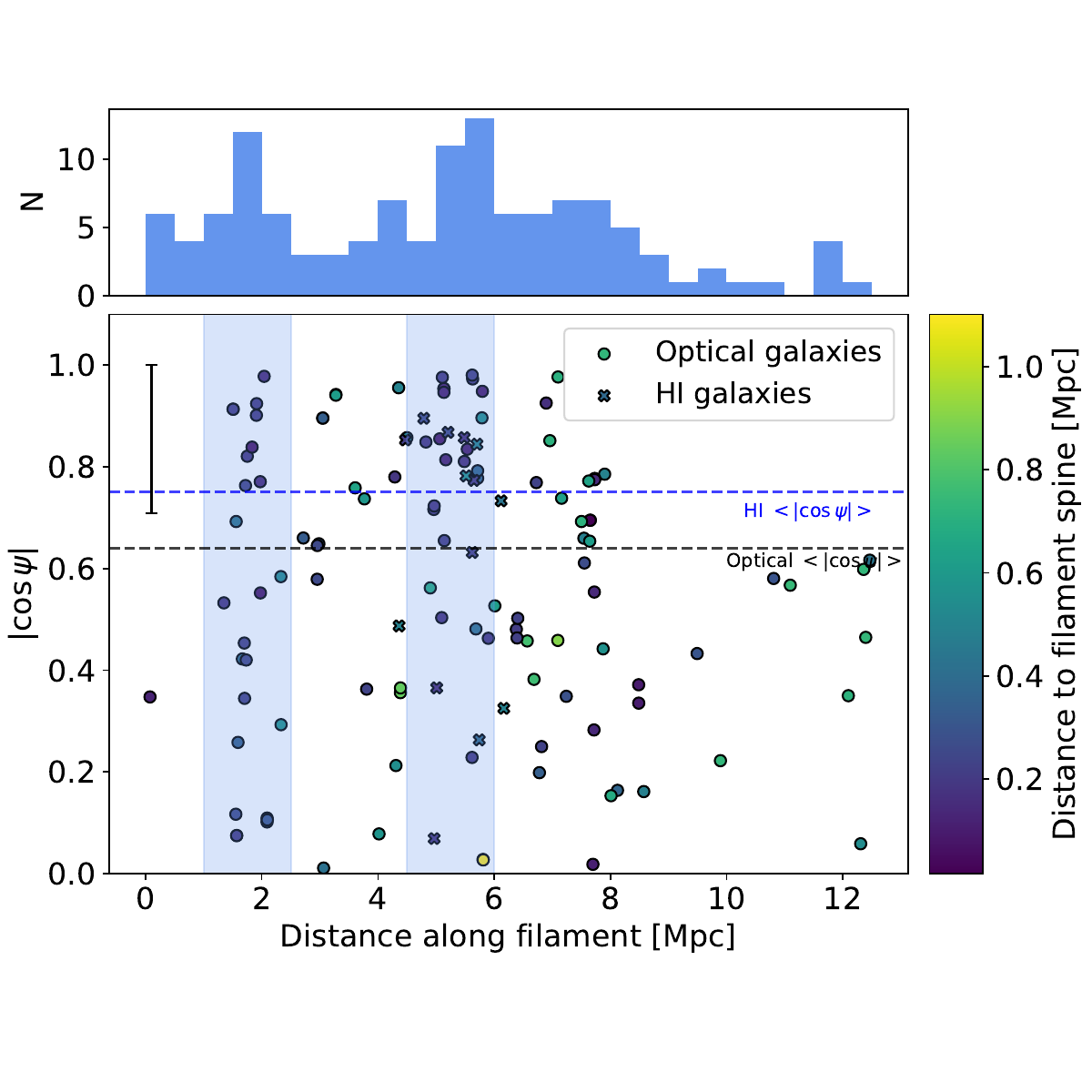} 
\caption{The galaxy-filament alignment ($\lvert \cos \psi \rvert$) for the galaxy sample as a function of projected distance along the filament for both the H{\sc i} galaxies (crosses) and optical galaxies (circles). The horizontal dotted line marks the median value of $\lvert \cos \psi \rvert$ (black, optical; blue, H{\sc i}). The colour bar shows the projected distance to the spine of the filament. The error bars (only one shown for clarity) are calculated by bootstrapping the filament $100$ times and by recalculating $\lvert \cos \psi \rvert$ each iteration. The top panel shows a histogram of the number of all galaxies (including the ones where we could not calculate the spin-filament alignment) in bins of $0.5$~Mpc along the spine of the filament from the bottom to the top and the shaded strips in the main panel show the two regions that have the highest galaxy density.}
\label{fig:filament-alignment}
\end{figure}

As gas supplied by cosmic filaments leaves an imprint on galaxy kinematics, theory predicts and simulations show that the spin axis of the galaxy will preferentially align with its closest filament \citep{Pichon_2011}, especially at distances within $\sim 1$~Mpc of the filament spine. As such, we compute the spin-filament alignment (the cosine of the angle; $\langle\lvert \cos \psi \rvert\rangle$) between the spin axis of each of the galaxies and orientation of the filament for both the H{\sc i} sample, using the dynamical information from the H{\sc i} spectral cubes, and the optical galaxies using the DESI Legacy Imaging (with the assumption that the major axis is perpendicular to the spin axis), where we measure the orientation and ellipticity of the galaxies to determine the vector corresponding to the minor axis of the optical galaxies.
Following convention \citep{Kraljic_2021, tudorache2022}, we define galaxies with $\lvert \cos \psi \rvert > 0.5$ as being aligned and $\lvert \cos \psi \rvert < 0.5$ as being misaligned. In Figure~\ref{fig:filament-alignment}, we show the alignment (or misalignment) as a function of distance along the spine of the filament.
For the $14$ galaxies in the H{\sc i} sample, we find that a median value $\langle\lvert \cos \psi \rvert\rangle = 0.75 \pm 0.05 $. For the optical sample, we find a median value of $\langle\lvert \cos \psi \rvert \rangle = 0.64 \pm 0.02$, indicating a similar preference for alignment. To also account for the error in the position of the filament itself for the spin-filament alignment, we recalculate $\langle\lvert \cos \psi \rvert\rangle$ for each new filament iteration. Therefore, for each galaxy, we had $101$ values for $\lvert \cos \psi \rvert$, which allows us to determine the standard deviation on each $\lvert \cos \psi \rvert$ value shown in Figure~\ref{fig:filament-alignment}.

\begin{figure*}
    \centering
\begin{subfigure}[b]{0.45\linewidth}
  \centering
  \captionsetup{justification=centering}
  \includegraphics[width=1\linewidth]{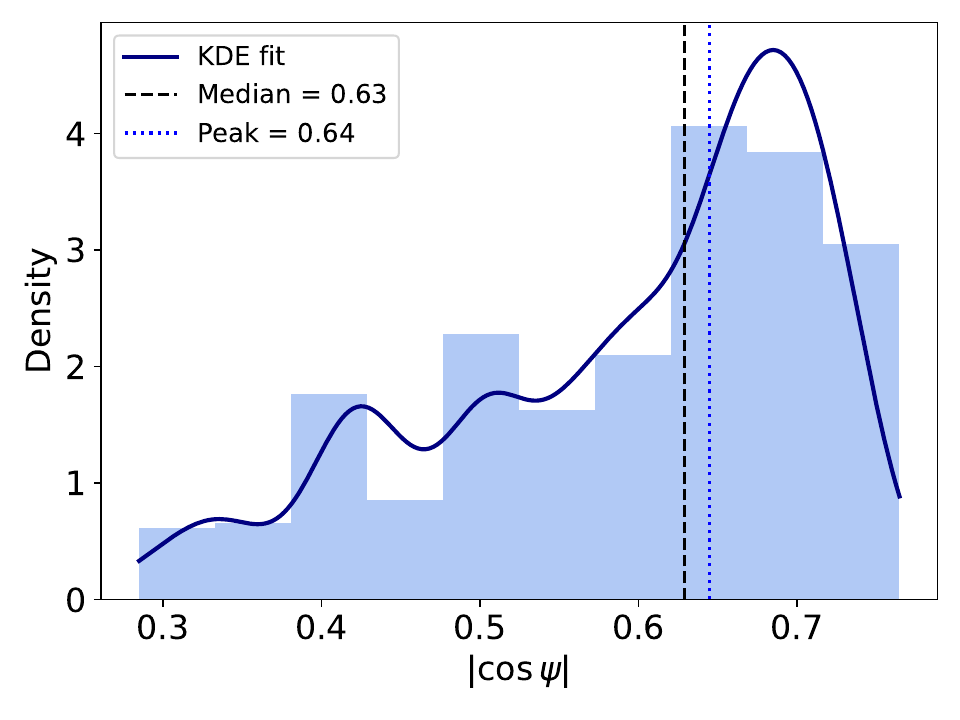} 
  \phantomsubcaption
  \label{subfig:hi-median-flip}
\end{subfigure}
\begin{subfigure}[b]{0.45\linewidth}
  \centering
  \captionsetup{justification=centering}
  \includegraphics[width=1\linewidth]{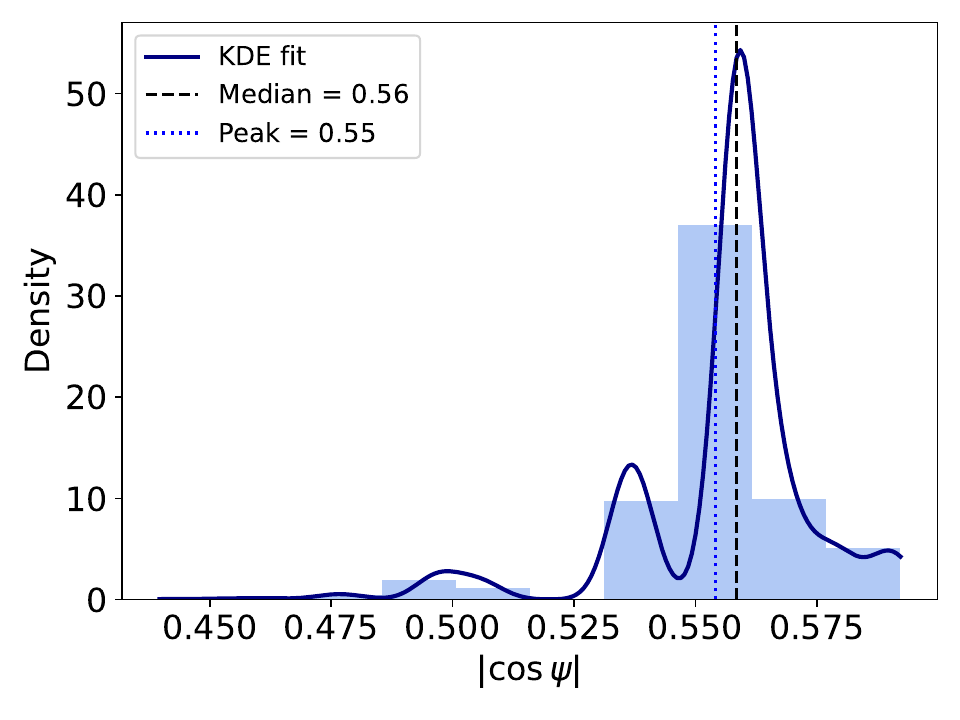} 
  \phantomsubcaption
  \label{subfig:optical-median-flip}
\end{subfigure}
    \caption{The distribution of $\langle\lvert \cos \psi \rvert \rangle$ for the optical sample ({\it left}) and the H{\sc i} sample ({\it right}) for the $2000$~iterations randomly assign the direction of measurement of the spin-axis/minor axis and recalculating the spin-filament distribution.}
  \label{fig:median-flip}
\end{figure*}

As mentioned in Section~\ref{subsec:spin-filament-calc}, there is also an uncertainty in the tilting of the galaxy, which introduces a degeneracy of $\pm \pi$ for the position angle of the galaxies. In order to quantify this effect, for each galaxy, we randomly assign the direction of measurement of the spin-axis/minor axis and recalculate the spin-filament alignment. We do this $2000$ times for each galaxy and then calculate the median of $\cos \psi \rvert \rangle$ for each iteration, obtaining $2000$ measurements of the median (mis-)alignment. Figure~\ref{fig:median-flip} shows the distribution of medians across all iterations. As can be seen, we find a peak of $\langle\lvert \cos \psi \rvert \rangle = 0.64 \pm 0.05$ for the  H{\sc i} galaxies and a peak of $\langle\lvert \cos \psi \rvert \rangle = 0.55 \pm 0.04$ for the optical galaxies. 
Therefore, after accounting for the uncertainty in the tilt of each galaxy, we find strong evidence for the spin axes (minor axes) to be aligned with the filament.

The degree to which we find the galaxies aligned with the filament is significantly higher than any current simulations predict. 
Using the SIMBA simulation \citep{simba-sim}, only marginal alignments were found between the spin-axis of H{\sc i}-rich galaxies and their cosmic filaments \citep{Kraljic_2020}, with a transition from $\langle\lvert \cos \psi \rvert\rangle \simeq 0.49$ for low-mass H{\sc i} galaxies ($M_{\text{H{\sc i}}} < 10^{9.5}$\,M$_{\odot}$) to a $\langle\lvert \cos \psi \rvert\rangle \simeq 0.51$ for high-mass galaxies ($M_{\text{H{\sc i}}} > 10^{9.5}$\,M$_{\odot}$)

The same study also found a transition from aligned to mis-aligned dependent on the stellar mass, with similar marginally different values: $\langle\lvert \cos \psi \rvert\rangle \simeq 0.505 \pm 0.002$ for low stellar mass galaxies ($10^{9.0}$\,M$_{\odot} < M_{\ast} < 10^{9.5}$\,M$_{\odot}$) to a $\langle\lvert \cos \psi \rvert\rangle \simeq 0.473 \pm 0.022$ for high stellar mass galaxies ($M_{\ast} > 10^{11.5}$\,M$_{\odot} $), with the transition occurring at $M_{\ast} = 10^{10.1 \pm 0.5}$\,M$_{\odot}$ at $z=0$. A similar transition is found in \cite{Codis_2018}, at a stellar mass of $M_{\ast} = 10^{10.1 \pm 0.3}$\,M$_{\odot}$, also with small $\langle\lvert \cos \psi \rvert\rangle$ variations. 

Whilst the stellar masses of our H{\sc i} galaxy sample all fall below this threshold and are preferentially aligned, the signal is significantly stronger than found in current simulations. This result is reinforced with the entire optical sample, where the degree of alignment that we find is similarly high in significance, irrespective of the stellar mass thresholds. 

It has been shown \citep{Codis_2015, Codis_2018, Malavasi_2022} that the alignment can depend on the density within filaments. This is expected, since as densities increase, galaxies interact more (via mergers) and the massive galaxies become tidally aligned. In Figure~\ref{fig:filament-alignment} we show that the fraction of galaxies that are misaligned depends on the galaxy number density, i.e. we find that misalignment is more prevalent in the areas where there is an over-density of galaxies. For example, if we split the sample where we see an overdensity of galaxies at  $4.5 < \mathrm{d_{\mathrm{fil}}}  < 6.0$\,Mpc along the filament spine, we find significantly higher fraction of galaxies with $\langle\lvert \cos \psi \rvert\rangle < 0.5$ ($ \approx 25$\%) compared to a fraction of 9\% for the neighbouring distance bin ($2.5 < \mathrm{d_{\mathrm{fil}}}  < 4.5$\,Mpc) of the filament. This provides some evidence for a disruption in the alignment between the orientation of galaxies and the large-scale filamentary structure in the densest regions. However, we do not see the same effect in the highest distance bin along the filament ($\mathrm{d_{\mathrm{fil}}}  > 8.0$\,Mpc)

\subsection{Filament rotation}
\label{subsec:rotation}

The relative velocities of the galaxies with respect to the filament can be used to determine whether the filament itself is rotating. This can provide information as to whether the filament itself could be a source of angular momentum that may cause the strong alignment between the spin axes of the galaxies and the filament.

In Figure~\ref{fig:rotation}, we show the position of the galaxies along the filament with the colour bar denoting their velocity relative to the filament spine. We can see that galaxies lying to the west of the filament spine tend to have positive velocities (i.e. receding), whereas those galaxies to the east of the filament spine have negative velocities (approaching). Whilst this is not necessarily definite proof of bulk rotation, since the redshift gradient can be interpreted as the structure being a cosmic wall, we pursue this avenue for the rest of this section.

To quantify how dominant the bulk rotation of galaxies in the filament is compared to the random motion, we calculate the ratio between the two velocity components, which can be interpreted as the dynamical temperature, $T_{\rm d}$ of the filament \citep{Wang_2021}:
\begin{equation}
T_{\rm d}  = \sigma_{\rm z}/\Delta z_{\rm AB} \approx1.235,
\end{equation}
where $\sigma_{\rm z}$ is the root-mean-square of the redshift distribution of all the galaxies with respect to the filament and $\Delta z_{\rm AB} = \langle z_{\rm A} \rangle - \langle z_{\rm B} \rangle$, where $\langle z_{\rm A} \rangle$ is the mean redshift of all the galaxies on the receding side and $\langle z_{\rm B} \rangle$ is the mean redshift of all the galaxies on the approaching side. 

\begin{figure}
  \centering
  \captionsetup{justification=centering}
  \includegraphics[width=1\linewidth]{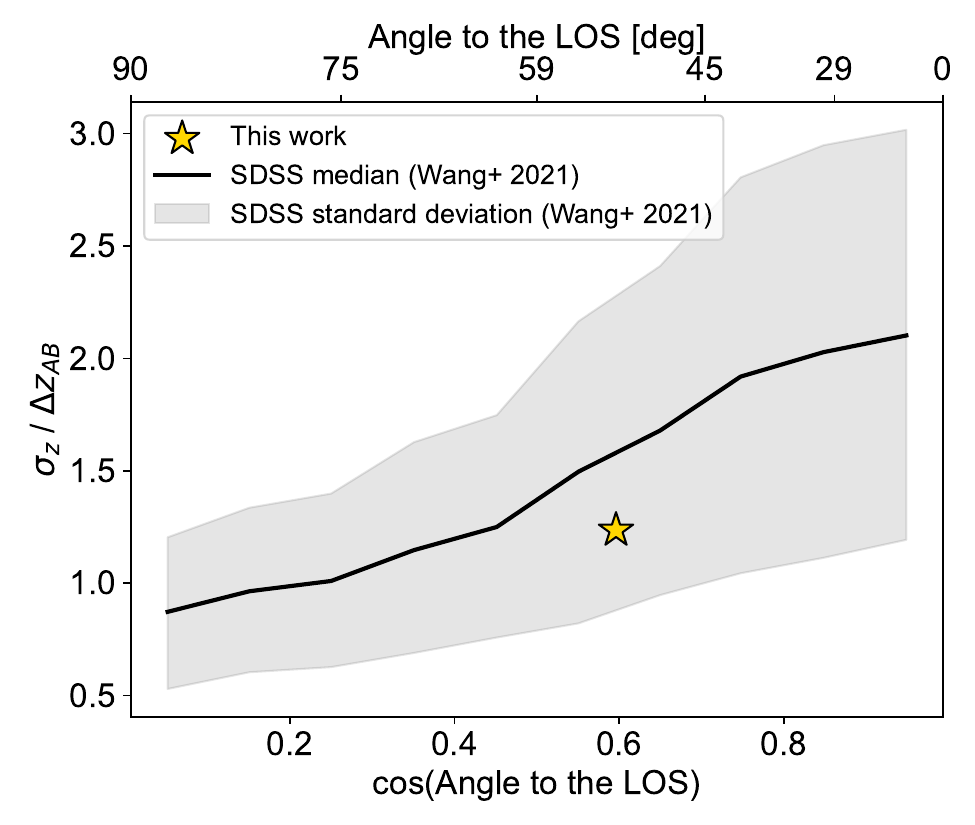} 
  \caption{The median dynamical "temperature" \citep[see][and text]{Wang_2021} of filaments as a function of the inclination angle between the filament spine and the line of sight for the filament (star) in comparison with the median calculated from the stacked filaments in SDSS (black line, with a standard deviation shown by the shaded region).}
\label{fig:filament-deltaz}
\end{figure}

This filament therefore has a lower dynamical temperature, i.e., more rotation-dominated, compared to the average of the cosmic filaments identified by the Sloan Digital Sky Survey \citep[SDSS;][\citealt{Wang_2021}]{sdss}, as can be seen in Figure \ref{fig:filament-deltaz}.

We subsequently model the rotation around the central spine assuming that the filament can be approximated as a pseudo-isothermal cylinder and describe the velocity as a function of radius:

\begin{equation}\label{eqn:rc}
v(R)=\sqrt{4 \pi G \rho_0 R_c^2\left[1 -  \frac{R_c}{R}\arctan{\frac{R}{R_c}} \right]},
\end{equation}
where $G$ is the gravitational constant, $\rho_0$ is the central density of the filament, and $R_C$ the core radius of the filament.

We fit for the rotation of the filament using Equation~\ref{eqn:rc}, leaving the core radius ($R_C$) and the central density of the filament ($\rho_0$) as free parameters. We use \textit{Regression and Optimisation with X and Y errors} \citep[\textsc{roxy};][]{Bartlett_2023}, which is a Hamiltonian Monte Carlo based method. We adopt a conservative uncertainty on both the velocity with respect to the filament spine and the distance of the galaxy from the spine, noting that we cannot measure the latter directly due to the fact that we only see the filament in projection. We adopt a 20 per cent uncertainty on the relative velocity and we jack-knife resample the galaxies to re-determine the filament each time using \textsc{Disperse}. This allows us to estimate the uncertainty on the projected distance of the galaxies from the $1000$ realisations of the filament spine. We then take the $16$th and $84$th percentiles of the distribution of the distance of each galaxy from the filament spine as the uncertainty on the distance.

\begin{figure*}
    \centering
\begin{subfigure}[b]{0.45\linewidth}
  \centering
  \captionsetup{justification=centering}
  \includegraphics[width=1\linewidth]{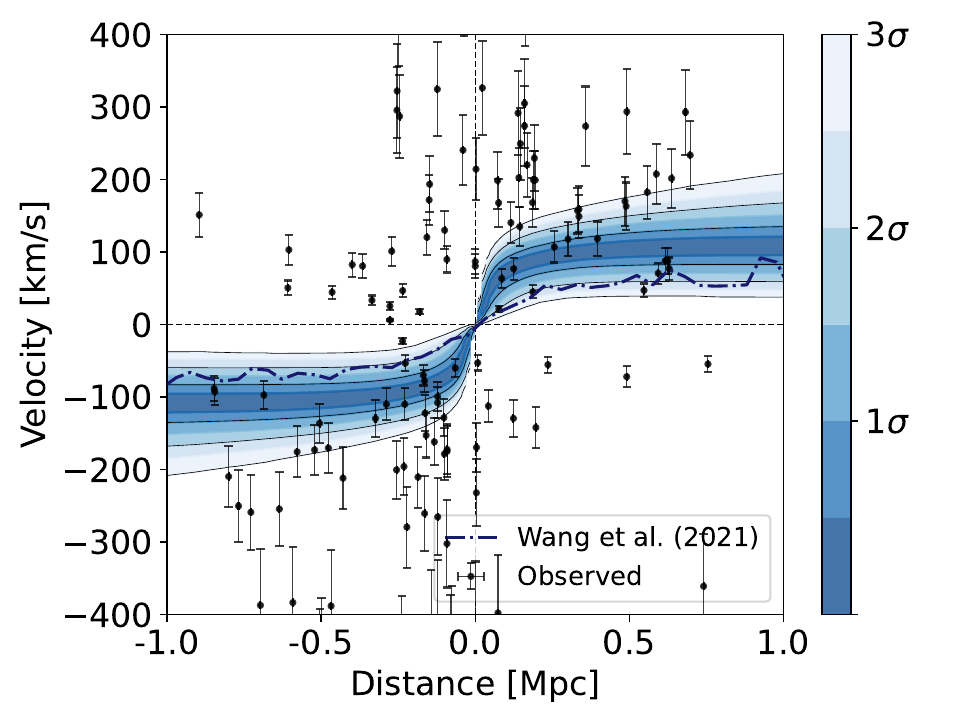} 
  \phantomsubcaption
  \label{subfig:roxy-curve-all}
\end{subfigure}
\begin{subfigure}[b]{0.45\linewidth}
  \centering
  \captionsetup{justification=centering}
  \includegraphics[width=1\linewidth]{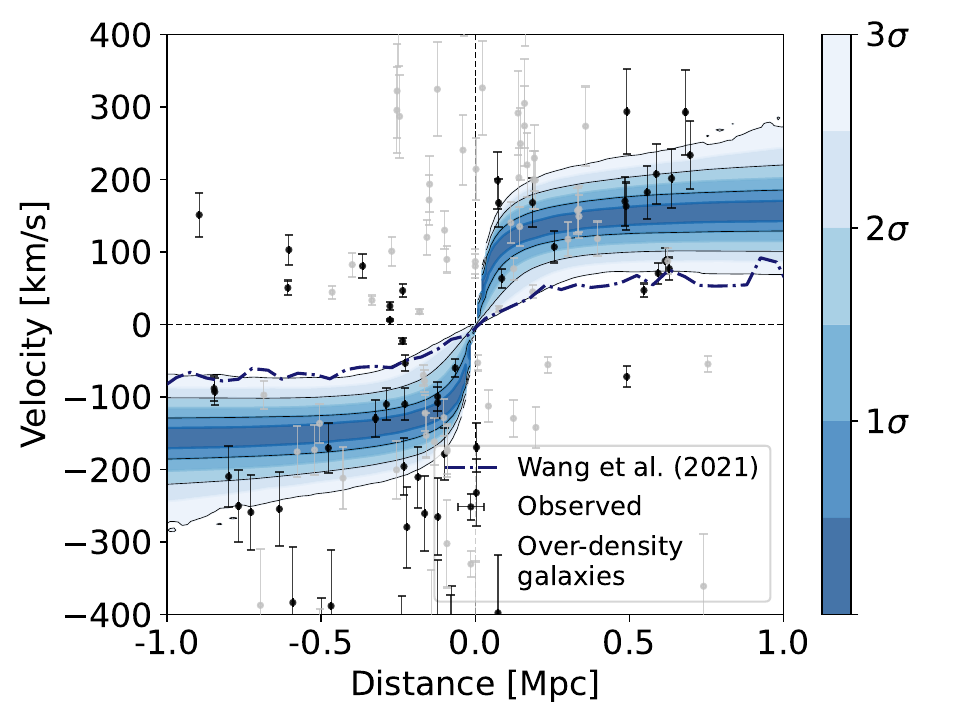} 
  \phantomsubcaption
  \label{subfig:roxy-curve-no-overdensity}
\end{subfigure}
    \caption{Projected distance from filament as a function of relative velocity of galaxies to the filament, for the whole sample ({\it left}) and for the sample without the galaxies located in the overdensities ({\it right}). The blue line is the fit of the rotation curve (see text) on the observed points, with the $2\sigma$ and $3\sigma$ regions shown on the colourbar. The black dot-dashed line is the average rotation curve using the large sample of filaments from SDSS.}
  \label{fig:d-v-mean-fil}
\end{figure*}

We obtain $\log_{10}(R_C /$\,kpc)$ = 1.72 \pm 0.36$ and $\log_{10}({\rho_0 /}$ M$_{\odot}~\mathrm{kpc}^{-3}) = 4.96 \pm 0.59$, which gives $\rho_0 = 3.49 \times 10^{-27}$~g~cm$^{-3}$. However, in the densest regions there is likely to be more random motion, evidenced by the increase in mis-alignments shown above, and such disruptions would also likely increase the velocity dispersion in these regions. We therefore fit the rotation curve to a sample in which we remove galaxies which reside in the most dense regions of the filament, $1.0~{\rm Mpc} < d_{\mathrm{fil}} < 2.5~$Mpc and $4.5~{\rm Mpc} < d_{\mathrm{fil}} < 6.0~$Mpc (shown in Figure \ref{fig:d-v-mean-fil}). For this fit, we obtain $\log_{10}(R_C /$\,kpc)$ = 1.7 \pm 0.29$ and $\log_{10}({\rho_0 /}$ M$_{\odot}~\mathrm{kpc}^{-3}) = 5.31 \pm 0.5$, which gives $\rho_0 \approx 1.39 \times 10^{-26}$~g~cm$^{-3}$.

The value for the core radius in simulations has been found to be of the order 5 -- 22~kpc \citep{Ramsoy_2021,Lu_2024} at $z>3$ and we find $R_C \sim 400$~kpc at $z\sim 0.032$. Whilst much larger than that found at high redshift in simulations, it matches the predicted trend in which the core radius of the filament increases as redshift decreases.
Conversely, we find that the core density, $\rho_0$ is lower than in the simulations at $z \sim 3-4$ ($\rho_{0} \sim 10^{-26.5} - 10^{-25.5} \text{g~cm}^{-3}$ in \citealt{Lu_2024} or $\rho_0 = 10^{-25.83 \pm 0.49} \text{g~cm}^{-3}$ in \citealt{Ramsoy_2021}). However, this is consistent with the picture of the core radius increasing towards lower redshift and consequently the density decreasing. Indeed, if we assume the redshift evolution of the core radius \citep{Ramsoy_2021} has the form $R_C \propto (1+z)^{-3.18}$, we would expect the filament at $z\sim 0.03$ to have a core radius of $R_C \sim 500$\,kpc, which is significantly higher than our measurement, although the extrapolation from high redshift is large and therefore likely within the range of low-redshift filamentary structures. As the core radius expands, we also expect the central density to decrease. For an isothermal
cylinder in hydrostatic equilibrium the central density $\rho_0 \propto R_C^{-1/2}$ \citep{Lu_2024}, i.e. a factor of $\sim 10$ lower density, which is also consistent (within the large uncertainties) with our measurement for $\rho_0$.
The value for the core radius from the dynamical measurements is much lower the characteristic radius determined using the galaxy number density and the derivative of the slope as defined in Section~\ref{sec:numdens}. However, these are measuring two different radii and should not be seen at the same characteristic scale.

To compare with simulations at lower redshift, we return to Figure~\ref{fig:filament-density-fit}, where we showed the galaxy density profile from the filament spine, alongside the derivative that quantifies how fast the density decreases with radius, following \citep{Diemer_2014} who use a similar method to study radial density gradients of dark matter halos in simulations. Our value for the effective radius from the galaxy number density of the filament fall within the range of those measured from the stacked filaments using SDSS ($R = 0.81$~Mpc) and those derived from the Millenium-TNG simulation ($R = 0.98$~Mpc) \citep{Wang_2024} using a similar analysis. 
Thus, although the rotation of the filament suggests a relatively low-density filament in terms of the total matter contributing to the gravitational potential, the galaxy density distribution is consistent with the typical filaments found in both observations and simulations at low redshift.

\section{Discussion}
\label{sec:discussion}

The filament discovered in this work is the third such structure discovered using H{\sc i} galaxies reported in the literature.  It is therefore likely that more such structures will be detected with the advent of newer, more powerful radio interferometric surveys (see \citealt{Lawrie_2025} for a prediction for MeerKAT from the \textsc{Simba} simulation). However, it is different to the one found by \cite{Arabsalmani_2025} as it is approximately three times longer in length and composed of fifteen times larger number of spectroscopically confirmed member galaxies. Similarly, the filamentary-like structure in \cite{Lawrie_2025} is technically more similar to an H{\sc i} group than a part of a cosmic filament. Furthermore, this remains the first singular structure of its kind to have this level of alignment and the possible presence of angular momentum along the spine.

The discovery of a single filament with a very high degree of galaxy spin axis alignment with the cosmic filament, alongside the bulk rotation of the filament is consistent with the picture of Tidal Torque Theory \citep[TTT,][]{Pichon_2011} followed by gravitational collapse. In this picture, proto-filaments in the early Universe acquired angular momentum due to asymmetries in the gravitational tidal field \citep{White_1984, Schafer_2009}. This angular momentum is conserved during gravitational collapse, forming elongated, rotating structures that align with the large-scale cosmic web. The filaments then acquire spin due to the angular momentum seeded by tidal fields and amplified through accretion.

Since filaments and their neighbouring galaxies or halos emerge from the same large-scale environment, they are expected to share a similar large-scale spin field \citep{Wang_2021, Xia_2021}. This shared origin can explain the observed alignment between the spin of galaxies and that of their nearby filaments. Furthermore, for 
the spin-filament alignment, not only is this consistent with TTT and with other observational studies using H{\sc i} and optically-selected samples \citep{bird2019chiles, tudorache2022}, it is by far the strongest preference for alignment seen thus far. The gas supplied by cosmic filaments will leave an imprint on galaxy kinematics, since it comes from the converging flow of a cosmic wall (i.e., the gas flows perpendicular to the filament orientation) that forms elongated filaments, affecting the galaxies within \citep{Xia_2021}. Hence, the spin axis of the galaxy will align with its closest filament at higher redshifts \citep{Cadiou_2022}, especially at such close distances. In this scenario, it is then expected that the gas will fall into galaxies via cold flows, feeding discs with angular-momentum rich gas which will be tidally aligned with the cosmic web \citep{Codis_2015}.

Another prediction \citep{Wang_2021} which is consistent with our study is that in denser environments, such as near galaxy clusters or nodes, filaments experience stronger tidal fields, which will amplify their angular momentum and lead to faster rotation of the filament. In contrast, filaments in low-density regions will exhibit weaker angular momentum growth due to reduced tidal interactions. We identify a plausible node along the spine of the filament (around $\sim 2$~Mpc along the spine of our defined filament), which has a large number of galaxies that have their spin-axes mis-aligned with the filament, which may act to increase the rotation of the filament. 
Furthermore, similar to the spin-filament alignment result, we clearly find more evidence for rotation of the filament for the galaxies outside the more dense regions. This is consistent with the fact that the interactions (tidal, via mergers) between these galaxies can disrupt the overall imprint of angular momentum from the large-scale structure.

We find that the alignment between the spin axes of the galaxies and the filament is significantly stronger than predicted by simulations \citep{Kraljic_2020} or found in previous observations \citep{Welker_2020}. 
If such strong alignments are common in the Universe, then this has the potential to contaminate weak lensing measurements by introducing an artificial correlation between background galaxies and foreground structures, effectively reducing the statistical precision of lensing-based cosmological constraints \citep{Chisari_2017}. As a result, if not properly accounted for, these alignments can bias weak lensing-derived constraints on the matter power spectrum and the growth rate of cosmic structures. Given the accuracy of upcoming surveys such as Euclid \citep{Laureijs_2011, Euclid_2024} and Vera C. Rubin Observatory \citep{LSST_2009, Ivezic_2019}, this may be a more significant contaminant that has been considered thus far.

\section{Summary}
\label{sec:conclusions}

In this work, we present the detection of $14$~H{\sc i} galaxies in the MIGHTEE-H{\sc i} survey at $z = 0.03$ in the COSMOS field which form a filamentary-like structure with the length of $1.7$~Mpc, width of $\sim 36$~kpc and have a velocity dispersion of $140$~km/s. We find that all the H{\sc i} galaxies are within $1.0$~Mpc of a cosmic web filament computed from SDSS and DESI galaxies. The DESI filament itself has a length of $15.4$~Mpc and a thickness between $\sim 0.8$~Mpc (when fitting a polynomial) and $\sim 1$~Mpc (when assuming it is a cylinder).

We calculate the spin-filament alignment $\lvert \cos \psi \rvert$ for each of the galaxies, and we find that there is a strong preference for the galaxies to be aligned with the cosmic web filament, as the median value of the whole sample is $\langle\lvert \cos \psi \rvert\rangle = 0.75 \pm 0.05$. Even when accounting for the uncertainty in the orientation of the position angle, we still obtain a statistically significant signal for alignment ($\langle\lvert \cos \psi \rvert\rangle = 0.64 \pm 0.05$). Similarly, when we repeat this for the optical galaxies from the DESI galaxies, we obtain a similar result (albeit weaker) for the alignment, with a median value of $\langle\lvert \cos \psi \rvert\rangle = 0.55 \pm 0.04$, when accounting for the unknown tilt direction.

Furthermore, we investigate the bulk motion of the optical galaxies within this filament. We find that the galaxies exhibit strong evidence for rotation around the spine of the filament - making this the longest spinning structure thus far discovered.

This structure therefore shows that within a cosmic filament, the H{\sc i} gas is relatively undisturbed in its angular momentum. As cosmological simulations predict the H{\sc i} gas will be feasible to detect with the rise of next-generation radio telescopes \citep{Kooistra_2017, Kooistra_2019}, this structure can prove to be the ideal environment to attempt such a detection.
By understanding the relationship between this H{\sc i} structure and the cosmic filament it traces, it has the potential to pin down the relationship between the low density gas in the cosmic web and how the galaxies that lie within it grow using its material.

\section*{Acknowledgements}

MNT, SLJ, MJJ, IH, AV and TY acknowledge the support of a UKRI Frontiers Research Grant [EP/X026639/1], which was selected by the European Research Council, and the STFC consolidated grants [ST/S000488/1] and [ST/W000903/1]. MNT, MJJ, AAP and IH also acknowledge support from the Oxford Hintze Centre for Astrophysical Surveys which is funded through generous support from the Hintze Family Charitable Foundation. MG is supported by the UK STFC Grant ST/Y001117/1. MG acknowledges support from the Inter-University Institute for Data Intensive Astronomy (IDIA). 
The MeerKAT telescope is operated by the South African Radio Astronomy Observatory, which is a facility of the National Research Foundation, an agency of the Department of Science and Innovation. We acknowledge use of the Inter-University Institute for Data Intensive Astronomy (IDIA) data intensive research cloud for data processing. IDIA is a South African university partnership involving the University of Cape Town, the University of Pretoria and the University of the Western Cape. The authors acknowledge the Centre for High Performance Computing (CHPC), South Africa, for providing computational resources to this research project. For the purpose of open access, the authors have applied a Creative Commons Attribution (CC BY) licence to any Author Accepted Manuscript version arising from this submission.

This research made use of Astropy,\footnote{\url{http://www.astropy.org}} a community-developed core Python package for Astronomy \citep{astropy:2013, astropy:2018}.

\section*{Data Availability}

The MIGHTEE-H{\sc i} spectral cubes are released as part of the first data release of the MIGHTEE survey, which include cubelets of the sources discussed in this paper \citep{Heywood_2022}. The derived quantities from the multi-wavelength ancillary data were released with the final data release of the VIDEO survey mid 2021. Alternative products are already available from the Deep Extragalactic VIsible Legacy Survey \citep[DEVILS;][]{devils}.



\bibliographystyle{mnras}
\bibliography{ref} 







\bsp	
\label{lastpage}
\end{document}